\documentclass[elsart]{elsart}
\usepackage{graphicx,amssymb,amsmath,times}
\journal{Astroparticle Physics}
\begin{document}
\begin{frontmatter}
\title{High energy diffuse gamma-ray emission of the galactic disk and Galactic Cosmic-Ray spectra}
\author{Thoudam Satyendra}
\address{Astrophysical Sciences Division, Bhabha Atomic Research Centre\\ Mumbai-400085, India}
\ead{satyend@barc.ernet.in}

\begin{abstract}

Observations of diffuse Galactic $\gamma$-ray spectrum by the EGRET instrument reveal an excess above $\sim$ 1 GeV over the expected $\gamma$-ray spectrum calculated under the assumption that the locally observed cosmic-ray (CR) spectra represent the galactic CR spectra. Assuming that Galactic CRs of energy below $\sim 100$ TeV are accelerated by supernova remnant (SNR) shock waves and that the shock compression ratio is SNR age dependent, the average source injection spectra from an ensemble of SNRs is calculated both in the inner ($330^\circ<l<30^\circ$) and outer ($30^\circ<l<330^\circ$) regions of the galaxy. The calculation considers the SNR age distribution in the galaxy. Injecting these spectra in the galaxy and using a 3-D convection-diffusion equation, the CR electrons and protons spectra in the two galactic regions are obtained and their spectra in the galactic disk are found to be flatter than the observed CR spectra. The diffuse gamma-ray spectrum produced by the interaction of these galactic CRs with the ISM and ISRFs is compared with the experimental data in both the galactic regions. Furthermore, the steepening of the observed local CR spectra from the galactic disk CR spectra are discussed by propagating local CRs having a source spectrum derived using local SNR age distribution (SNRs located within $1.5kpc$ from the Sun), for a diffusion coefficient $D_0\sim 0.3\times 10^{27} cm^2 s^{-1}$ in the local region which is much less than the typical value in the galaxy $D_0\sim (1-10)\times 10^{28} cm^2 s^{-1}$. The results obtained in this paper support the SNR origin of galactic CRs.
\end{abstract}
\end{frontmatter}

\section{Introduction}
The spectrum of radiation produced by the interactions of galactic CRs with the interstellar medium (ISM) and the interstellar radiation field (ISRF) gives information about their propagation in the galaxy, their sources distribution and their initial source spectrum. Results from observations made with the Energetic Gamma Ray Experiment Telescope (EGRET) on the Compton Gamma-Ray Observatory have shown that above $\sim$ 1 GeV, the diffuse Galactic $\gamma$-ray spectrum exceeds significantly the spectrum calculated using the locally measured proton and electron spectra [1]. Explaining the excess with the $\pi^0$-decay gamma-ray alone requires a proton spectrum with index $\sim$ 2.4$-$2.5 [2] which is much flatter than $\sim$ 2.7, the locally measured value [3]. Many authors have given different views to explain the excess (see e.g. [4-9]), but, still the problem is not fully understood.

Although supernova explosions are considered to be the main sources of CRs with energies up to $\sim10^{14}$ eV in the galaxy (see e.g. [10]), most of the SNR statistics (except the radio spectral index distribution and the source density radial distribution) have not been used extensively in studying this puzzle of $``GeV excess"$ and the observed cosmic-ray spectra. In most of the earlier studies (e.g. [4,5]), the CR source injection spectrum was taken either from the radio observations (taken at a few $GH_z$ which corresponds to an electron energy $\sim 5 GeV$) or as an input parameter which was then optimized to the observed $\gamma$-ray data. In this paper, the injection spectrum is first calculated independent of the $\gamma$-ray data by considering shock acceleration mechanisms and then injected into the galaxy and see whether the excess can be explained. This is more realistic compared to the earlier studies. 

In spite of the fact that $\sim 77\%$ of the total 231 galactic SNRs detected in radio are shell(S)-type, statistical studies have found that there is no firm relationship between the radio spectral index and the age of the remnant [11,12] while ideally shock acceleration theory predicts a correlation between the particle spectral index $q(t)$ (which relates with the radio index) and the shock velocity (which in turn relates with the SNR age) as [13]
\begin{equation}
q(t) =\frac{3s(t)}{s(t)-1}=\frac{4M(t)^2}{M(t)^2-1}
\end{equation}
where $s(t)=\frac{4{M(t)}^2}{{M(t)}^2+3}$ is the shock compression ratio, $M(t)=u(t)/c_s$ is the Mach number of the shock defined as the ratio of the shock velocity $u(t)$ at an age $t$ to the sound speed of the unshocked intersteller medium $c_s$ and 
\begin{equation}
u(t)=u_0,\qquad t<t_{sed},\qquad \qquad \qquad \qquad \qquad \qquad \quad(1a)\nonumber\\
\end{equation} 
\begin{equation}
=u_0\left(\frac{t_{sed}}{t}\right)^{0.6},\qquad t_{sed}\leq t<t_{rad},\qquad \qquad \qquad \qquad(1b)\nonumber
\end{equation} 
\begin{equation}
=u_0\left(\frac{t_{sed}}{t_{rad}}\right)^{0.6}\left(\frac{t_{rad}}{t}\right)^{0.69},\qquad t_{rad}\leq t\qquad\qquad\qquad\quad (1c)\nonumber\\
\end{equation} 
$t_{sed}=(3M_{ej}/4\pi\mu m_H n_H u_0^{3})^{1/3}$ denotes the time which marks the start of the Sedov phase where $M_{ej}$ is the ejecta mass, $\mu=1.4$ is the mean atomic weight of the ISM, $m_H$ is the hydrogen mass, $\bar{n}_H$ is the mean ISM hydrogen density and $u_0=\sqrt{2E_{SN}/M_{ej}}$ is the initial shock velocity with $E_{SN}$ repesenting the energy released in SN explosion. $t_{rad}=2.7\times 10^4E_{51}^{0.24}n_H^{-0.52}$ yr is the start of the radiative phase which also represents the end of the Sedov phase where $E_{51}$ is the SN explosion energy in units of $10^{51}$ ergs [14]. The non-correlation between the radio index and the SNR age requires a very detailed study and will not be discussed here, but one possibility may be due to the presence of more than one electron population, as is well known in the case of Crab nebula. 

In section 2 an estimate of the SNR age distribution in the galaxy is given. This distribution is then used in section 3 for calculating the particle source spectra in the inner ($330^\circ<l<30^\circ$) and outer ($30^\circ<l<330^\circ$) regions of the galaxy. The resulting source spectra are further used to obtain the respective galactic CR spectra. In section 4, using the galactic CR spectra the diffuse gamma-ray flux is calculated. In section 5, the calculation of the local CR spectra is given, and in section 6, the over-all results and the limitations of the model are discussed briefly.  
\begin{figure}[h]
\centering
\includegraphics*[width=0.5\textwidth,angle=270,clip]{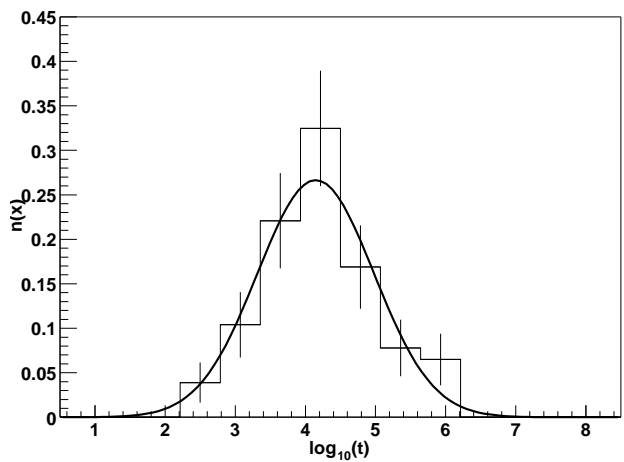}
\caption{\label {fig1} Normalised distribution of 77 S-type SNRs with known ages in the galaxy. The thick curve is the fitted gaussian (equation 2) with $a_0=0.266\pm 0.045$, $a_1=4.147\pm 0.104$ and $a_2=0.831\pm 0.113$} 
\end{figure}                                                 
\section{SNR age distribution}
Table 1 in Jian et al. 2005 [12] had given a list of several parameters of 231 all-type SNRs, out of which only 77 S-type SNRs are with known ages which have been estimated by different methods as indicated in the list. The normalized distribution of these 77 S-type SNRs (known ages) is shown in Fig.1. The distribution can be fitted with the gaussian 
\begin{equation}
 n(x)  = a_0 exp\left[-0.5\left({\frac{x-a_1}{a_2}}\right)^{2}\right]
\end{equation}
where $x=log_{10}t$ and $t$ is the SNR age in years. The fitted parameters together with their statistical errors, reduced chisquare ($\chi^{2}$/no. of degrees of freedom) and fit probability $(Prob)$ are given in Table 1. In the present study, it will be assumed that the age disrtribution in Fig.1 represents a galactic averaged S-type SNR age distribution in the galaxy. 

It can be argued that the age distribution shown in Fig.1 may not represent the true age distribution in the galaxy because of selection biases in radio observations which arises mainly due to the limited sensitivity, field of view and angular resolution of the instruments. Among these, only the limitation due to instrumental sensitivity is likely to affect the SNR age distribution strongly since it is directly related to the brightness of the source. Other effects are only likely to reduce the sample size as a whole irrespective of the SNR age. But, since the surface brightness decreases with SNR age, the sensitivity biasing will have their effects mostly on old SNRs. In fact, assuming adiabatic phase the surface brightness estimated from an SNR of age $\sim 10^5$ yrs is fainter than the detection limit in studies of Galactic SNRs distribution [15,16]. For the present study, we are mainly interested in SNRs with ages between approximately $10^3$ yrs and $3\times 10^4$ yrs (the start and end of the Sedov phase respectively, see next section) since particle acceleration is assumed to occur mainly in the Sedov phase of the SNR evolution. So, it will be assumed that the effects of detection biases due to the finite sensitivity of radio telescopes are quiet small in the present study. One more approximation concerns the confinement time of CRs in SNRs. Detailed treatment of particle acceleration in supernova shock waves involves the confinement of accelerated particles in the shock region until the shock becomes weak. Although the most energetic particles start escaping from the acceleration site already at the start of the Sedov phase, the main fraction of particles remain confined in SNRs for approximately ($10^4-10^5$) yrs [17]. But, for simplicity, it is assumed here that after acceleration (which starts at the onset of the Sedov phase) the particles start escaping rapidly from the shock region into the ISM. Therefore, from now on it will be assumed that the age distribution shown in Fig.1 also represents the distribution of CR acceleration times in the galaxy.

One interesting point that can also be noted is that the total spectrum from an ensemble of SNRs is dominated mainly by particles liberated from SNRs which are near the Sedov phase, except at the very low energy end of the spectrum (near particle injection energy $\sim$ 1 KeV). It is because of the high acceleration efficiencies of young SNRs which can give rise to flatter particle spectra in contrast to low acceleration efficiencies of old SNRs which give them a comparatively steeper source spectra. So, it can be concluded that almost the whole range of the total particle spectrum depends strongly on $t_{sed}$ and very weakly on $t_{rad}$. A typical hydrogen density of $n_H\sim 0.5 cm^{-3}$ gives $t_{sed}\sim 1.3\times 10^3$ yrs and $t_{rad}\sim 3.8\times 10^4$ yrs, both of which lie well within the detection limit of radio telescopes. Therefore, the effects of detection biases on the final source spectrum are definitely small. However, for a hydrogen density much lower than the typical galactic averaged value, say $n_H\sim 0.01 cm^{-3}$, $t_{sed}\sim 4.7\times 10^3$ and $t_{rad} \sim 3\times 10^5$. Here also, although $t_{rad}$ lies below the detection limit, the effect of the SNR selection biases on the total source spectrum will be considerably small as long as $t_{sed}$ lies well within the sensitivity of radio telescopes since young SNRs whose ages are around $t_{sed}$ dominate the final spectrum. But, such an extreme condition is far away from the typical galactic conditions ($n_H\sim 0.5 cm^{-3}$) which is considered in this work.  
 
\begin{table}   
\caption{Fit parameters for the normalized age distribution of 77 S-type SNRs with known ages shown in Fig.1 :}
\centering
\begin{tabular}{|c|c|} 
\hline
$Entries$ & 77           \\
\hline
$\chi^{2}/ndf$ & 3.258/4 \\
\hline
$Prob$		&  0.516     \\
\hline
$a_0$       &  0.266$\pm$0.045\\
\hline                                             
$a_1$       &  4.147$\pm$0.104\\
\hline                                                
$a_2$       &  0.831$\pm$0.113\\
\hline
\end{tabular}
\end{table}

\section{Galactic cosmic-ray spectrum}
\subsection{Source CR spectrum}
Considering the similarity of the power supplied by a SNe ($10^{51} erg/30yrs\sim 10^{42}erg/s$) to the power required to maintain the CR energy density in the galaxy ($\sim 10^{41}erg/s$), it is widely beleived that the majority of CRs upto $\sim 100 TeV$ are accelerated in SNRs. There is a wealth of literatures available which discusses the theory of diffusive shock acceleration both in a simple planar shocks model (see e.g. [18-20]) and in more detailed models that consider the shock goemetry as well as the non-linear CR backreaction effects (see e.g. [21-23]). In the present study, the non-linear effects are neglected and the much simpler planar shocks model will be adopted.
 
Acceleration of cosmic rays in plane, steady shocks in which the cosmic rays do not influence the shock structure  produces a power law spectrum of the form  
\begin{equation}
f(p,t) = q(t)p^{-q(t)}\int_0^p dp^{'}g(p^{'},t){p^{'}}^{q(t)-1}
\end{equation} 
where $f(p,t)$ is the isotropic accelerated particle distribution function with momentum $p$ arising from an injected particle distribution $g(p^{'},t)$ with momentum $p^{'}$ at time $t$. For constant injection of monoenergetic particles of momentum $p_0$ with density $n_0$,
\begin{equation}
g(p^{'},t)=\frac{n_0}{4\pi {p_0}^2}\delta(p-p_0)
\end{equation} 
which gives the accelerated particle spectrum at time $t$ as
\begin{equation}
f(p,t) = \frac{n_0}{4\pi p_0^3}q(t)(p/p_0)^{-q(t)}
\end{equation} 
Assuming that the SNR age distribution shown in Fig.1 exists in the galaxy at any point of time and injects CRs continuously in the galaxy, the particle source spectrum from an ensemble of SNRs is calculated by multiplying equation (5) by the SNR age distribution function (equation 2) and integrating from $t_0=t_{sed}$ (the start of the Sedov phase) to $t_f=t_{rad}$ (the end of the Sedov phase) as
\begin{equation}
f(p) = \frac{n_0}{4\pi p_0^3}\int^{t_f}_{t_0} q(t)(p/p_0)^{-q(t)}[n(x)/t]dt 
\end{equation} 
For calculating the actual particle source spectrum in the galaxy, the averaged ISM hydrogen atom  densities in the inner and outer regions of the galaxy are calculated as $\bar{n}_H=\int_{R_1}^{R_2}\bar{n}_H(R)dR/(R_2$-$R_1)$ where $R$ is the galactic radius, $\bar{n}_H(R)=\int_{0}^{h}n_H(R,z)dz/h$ and $n_H(R,z)=n_{HI}(R,z)+2n_{H_2}(R,z)$. Here, ($R_1,R_2$) are taken as $(2.25,8.5) kpc$ and $(5,16) kpc$ for calculations in the inner and outer galaxy respectively and $z$ represents the galactic height. The atomic hydrogen density distribution $n_{HI}(R,z)$ is taken from [24] and the molecular hydrogen density $n_{H_2}(R,z)$ is from [25]. Different authors had given slightly different values for the CO-to-H$_2$ conversion factor $X=N(H_2)/W(CO)$ where $N(H_2)$ is the molecular hydrogen column density and W(CO) is the velocity integrated CO intensity. $X=3.6\times 10^{20}$ $cm^{-2}$ (K km $s^{-1})^{-1}$ by Sanders et al.1984 [26], $X=(2.8\pm 0.7)\times 10^{20}$ $cm^{-2}$ (K km $s^{-1})^{-1}$ by Bloemen et al. 1986 [27] and $X=1.9\times 10^{20}$ $cm^{-2}$ (K km $s^{-1})^{-1}$ by Strong $\&$ Mattox 1996 [28]. Here, the value given by Strong $\&$ Mattox(1996) is adopted. Such a distribution of $n_{HI}(R,z)$ and $n_{H_2}(R,z)$ gives a mean hydrogen atom density of $\bar{n}_H\simeq 0.5(0.3)$ $cm^{-3}$ in the inner(outer) galaxy for $z\leq 700 pc$. Correspondingly, this gives a value of $t_{sed}=1.3\times 10^3$ yr $\&$ $t_{rad}=3.8\times 10^4$ yr in the inner region and $t_{sed}=1.5\times 10^3$ yr $\&$ $t_{rad}=5\times 10^4$ yr in the outer galaxy. Using these values and considering a total energy released of $E_{SN}=10^{51}$ ergs in SN explosion together with a total ejecta mass of $M_{ej}=8M_\circledS$ where $M_\circledS$ is the solar mass, the CR proton and electron source spectra can be calculated from equation (6) for the normalized age distribution shown in Fig. 1. The resulting energy spectra are plotted in Figs. 2$\&$3 for the inner $(330^\circ<l<30^\circ)$ and outer $(30^\circ<l<330^\circ)$ regions of the galaxy respectively for the case when particles are injected at energy 1 KeV with number density $n_0= 10^{-3}$ cm$^{-3}$. Detailed inspection of Figs. 2$\&$3 reveals the following:
\begin{figure}[h]
\centering
\includegraphics*[width=0.5\textwidth,angle=270,clip]{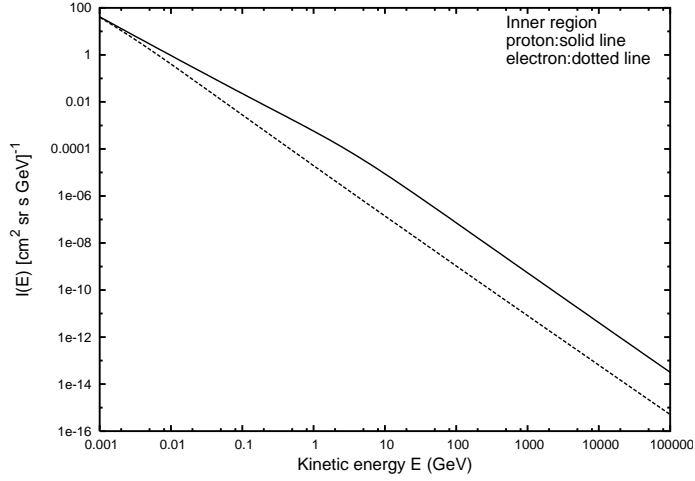}
\caption{\label {fig2} CR source spectra in the inner region ($330^\circ<l<30^\circ$) of the galaxy calculated assuming $E_{SN}=10^{51}$ ergs, $M_{ej}$= $8M_\circledS$, $\bar{n}_H=0.5 cm^{-3}$, $n_0= 10^{-3}$ cm$^{-3}$, injected particle energy of 1 KeV and an upper energy cut-off of $100$ TeV. Proton source spectrum (solid line): spectral index $\Gamma _p\simeq 2.12\pm 0.01$ for $E>>M_p$ and $1.66\pm 0.01$ for $E<<M_p$. Electron source spectrum (dashed line): $\Gamma _e\simeq 2.126\pm 0.004$}
\end{figure}
\begin{figure}[h]
\centering
\includegraphics*[width=0.5\textwidth,angle=270,clip]{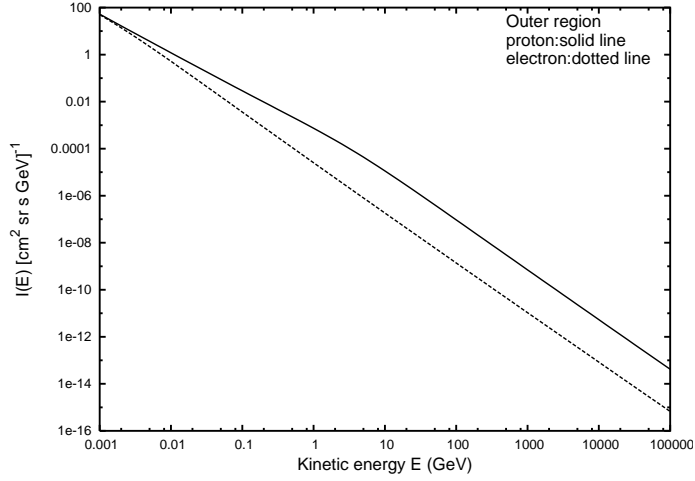}
\caption{\label {fig3} CR source spectra in the outer region ($30^\circ<l<330^\circ$) of the galaxy. All the parameters are same as in Fig. 2 except $\bar{n}_H=0.3 cm^{-3}$. Proton source spectrum (solid line): $\Gamma _p\simeq 2.11\pm 0.01$ for $E>>M_p$ and $1.66\pm 0.01$ for $E<<M_p$. Electron source spectrum (dotted line): $\Gamma _e\simeq 2.124\pm0.004$ }
\end{figure}

(i) For protons, the energy spectrum follow a broken power-law $I(E)\propto E^{-\Gamma _p}$ with a break at $\sim$ $M_p$, where $E$ is the kinetic energy and $M_p$ is the proton rest mass energy. The spectral index $\Gamma _p\simeq 2.12\pm 0.01$ for $E>>M_p$ and $1.66\pm 0.01$ for $E<<M_p$ for the inner galaxy and  $2.11\pm 0.01$ for $E>>M_p$ and $1.66\pm 0.01$ for $E<<M_p$ for the outer galaxy.

(ii) For electrons with $E>>M_e$ where $M_e$ is the electron rest mass energy, the spectrum has a power law index $\Gamma _e\simeq 2.126\pm 0.004$ for the inner region and $2.124\pm 0.004$ for the outer region.
\begin{figure}[h]
\centering
\includegraphics*[width=0.5\textwidth,angle=270,clip]{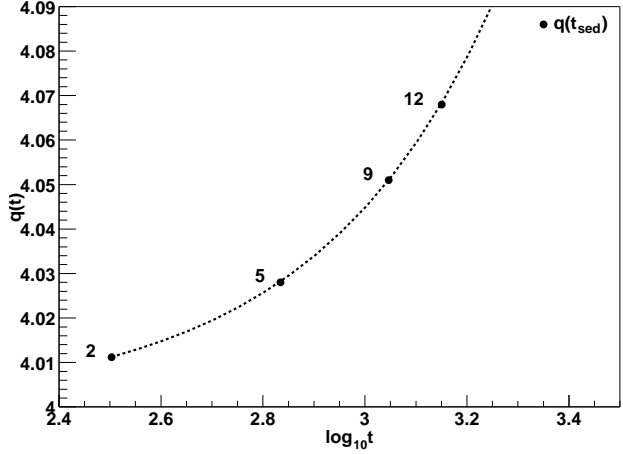}
\caption{\label {fig4} Plot of $q(t)$ vs $log_{10}t$ for the inner galaxy, where $t$ is the SNR age in years. The solid dots ($\bullet$) marked as 2,5,9$\&$12 are the values of $q(t_{sed})=(4.011,4.028,4.051,4.068)$ for $M_{ej}=(2,5,9,12)M_\circledS$ respectively. The sedov phase ends at $t_{rad}=3.8\times 10^4 yr$ which is not shown in the plot. The dashed curve is the particle momentum index q(t) obtained from equations (1) $\&$(1b) for SN explosion energy $E_{SN}=10^{51} ergs$.  }
\end{figure}

These spectra are almost consistent for a wide range of ejecta masses. For example: In the inner region, the spectral index changes from (2.091$-$2.136) for protons with $E>>M_p$ and from (2.102$-$2.146) for electrons, i.e. by a value of $\sim 0.044$ for ejecta masses $(2-12)M_\circledS$ for the same explosion energy of $10^{51}$ ergs. This mass range covers a large number of both the Type I and Type II supernovae. This consistency can be understand as follows: When the total spectrum from an ensemble of SNRs is calculated from equation (6), because of their high acceleration efficiencies which gives rise to a comparatively flatter spectra, those which are near the Sedov phase ($t_{sed}$) dominate the final spectrum except at the very low energy end of the spectrum (near particle injection energy $\sim$ 1 KeV) as mentioned in the last section. Fig. 4 shows the $q(t)$ vs $t$ plot for the sedov-phase obtained from equations (1) $\&$(1b). Since $u_0\propto M_{ej}^{-1/2}$ and $t_{sed}\propto (M_{ej}/u_0^3)^{1/3}$,  it is found that $t_{sed}$ increases with $M_{ej}$ as $t_{sed}\propto M_{ej}^{5/6}$. Also shown in the figure are the values of $q(t_{sed})=(4.011,4.028,4.051,4.068)$ for $M_{ej}=(2,5,9,12)M_\circledS$ respectively which changes by a small value of $\sim 0.057$ for the specified mass range. This weak dependence of $q(t_{sed})$ on $M_{ej}$ is expected since $q(t_{sed})=8E_{SN}/(2E_{SN}-c_s^2M_{ej})$ from equations (1) $\&$(1b). This slow variation of $q(t_{sed})$ is responsible for the small changes $\sim 0.044$ in the spectral index of the final particle spectrum obtained from equation (6) which is the superposition of individual spectra from various SNRs with ages ($t_{sed}\leq t\leq t_{rad}$).

It should be mentioned at this point of discussion that the mean hydrogen atom density obtained here may be less by a factor of upto $\sim 2$ than that used for calculating the diffuse gamma-rays from a particular direction in the galaxy because of the dominant emission of the radiation fluxes from a region very close to the galactic plane, say within a region of $z\lesssim 100 pc$ [6]. Thus, for calculating the diffuse gamma-ray fluxes later in section 4, an averaged atomic hydrogen density of $\sim 1(0.6) cm^{-3}$ will be adopted in the inner(outer) galaxy.

\subsection{Propagation of CR protons in the galaxy}
In the framework of a stationary 3D convection-diffusion model, the propagation of cosmic ray protons in the galaxy can be represented by the equation  
\begin{eqnarray}
\nabla\cdot \lbrace D(\textbf{r},E)\nabla N_p(\textbf{r},E)-\textbf{V}(\textbf{r})N_p(\textbf{r},E)\rbrace\qquad\qquad\qquad\qquad\qquad\qquad\nonumber\\
+\frac{\partial}{\partial E}\left\lbrace \left( \frac{\nabla\cdot\textbf{V}}{3}E-\beta_p(\textbf{r},E)\right) N_p(\textbf{r},E) \right\rbrace -\frac{N_p(\textbf{r},E)}{\tau_{pp}(\textbf{r},E)}+Q_p(\textbf{r},E)=0
\end{eqnarray}
where $E$ is the proton kinetic energy in GeV, $N_p(\textbf{r},E)$ is the differential number density , $D(\textbf{r},E)=D(E)=D_0(1+E/E_0)^\alpha$ [29] is the diffusion coefficient which is assumed to be independent of $\textbf{r}$ with $\alpha=0.6$, $E_0=3$ GeV and $D_0=(1-10)\times 10^{28} cm^2 s^{-1}$, $Q_p(\textbf{r},E)$ is the proton production rate, i.e.  $Q_p(\textbf{r},E)d^3rdE$ is the number of protons produced by the source in a volume element $d^3r$ in the energy range $(E,E+dE)$ per unit time, $\textbf{V}(\textbf{r})=3V_0z\hat{z}$ is the convection velocity in the direction perpendicular to the galactic plane with $V_0=15 km s^{-1} kpc^{-1}$ and $\beta_p(\textbf{r},E) = -B_p(E)$ is the proton energy loss rate due to ionization and Coulomb interactions. Following Mannheim and Schlickeiser 1994 [30], if $E$ is in GeV and proton mass energy, $m_p=0.938$ GeV, then 
\begin{eqnarray}
B_p(E)\approx 1.82\times 10^{-16} \bar{n}_H \left(\frac{E}{m_p}\right)^{-0.5}\left(1+0.85\frac{\bar{n}_{HII}}{\bar{n}_H}\right) GeV s^{-1}
\end{eqnarray} 
where $\bar{n}_{HII}$ is the mean density of ionised gas in cm$^{-3}$, $[-N_p(E,\textbf{r})/\tau_{pp}(E,\textbf{r})]$, $\tau_{pp}=E/(dE/dt)_\pi$ and ${(dE/dt)}_\pi=8\times 10^{-16} \bar{n}_H E$ in GeV s$^{-1}$ are the catastrophic loss term, energy loss time scale and  energy loss rate respectively due to pion production processes. With these substitutions, equation (7) can be simplified in rectangular coordinates $(x,y,z)$ as
\begin{eqnarray}
D(E)\left\lbrace \frac{\partial^2 N_p}{\partial^2 x}+\frac{\partial^2 N_p}{\partial^2 y}+\frac{\partial^2 N_p}{\partial^2 z}\right\rbrace-3V_0\frac{\partial}{\partial z}(zN_p)+\frac{\partial}{\partial E}\left\lbrace \left( V_0E-\beta_p(E)\right) N_p \right\rbrace\nonumber\\
-\frac{N_p}{\tau_{pp}}=-Q_p(\textbf{r},E)\qquad\
\end{eqnarray} 
The Green function of equation (9), i.e. the function $G(x,x^{'},y,y^{'},z,z^{'},E,E^{'})$ satisfying the equation 
\begin{eqnarray}
D(E)\left\lbrace \frac{\partial^2 G}{\partial^2 x}+\frac{\partial^2 G}{\partial^2 y}+\frac{\partial^2 G}{\partial^2 z}\right\rbrace-3V_0\frac{\partial}{\partial z}(zG)+\frac{\partial}{\partial E}\left\lbrace \left( V_0E-\beta_p(E)\right) G \right\rbrace\nonumber\\
-\frac{G}{\tau_{pp}}=-\delta(x-x^{'})\delta(y-y^{'})\delta(z-z^{'})\delta(E-E^{'})\qquad
\end{eqnarray} 
is found to be 
\begin{eqnarray}
G(x,x^{'},y,y^{'},z,z^{'},E,E^{'})=\frac{A_p(E)}{8\pi^{3/2}P_p(E)}{\left[\int^{E{'}}_E \frac{D(u)du}{P_p(u)}\right]}^{-1}\nonumber\\
\times{\left[\int^{E{'}}_E \frac{D(u)A_p^2(u)du}{P_p(u)}\right]}^{-1/2}exp\left[A_{pp}\int^E_{E^{'}} \frac{du}{P_p(u)}\right]\nonumber\\
\times exp\left[\frac{-((x^{'}-x)^2+(y^{'}-y)^2)}{4\int^{E^{'}}_E \frac{D(u)du}{P_p(u)}}\right]\nonumber\\
\times exp\left[\frac{-\left(3V_0z^{'}\int^E_{E^{'}}\frac{A_p(u)du}{A_p(E)P_p(u)}-z^{'}+z\right)^2}{4\int^{E^{'}}_E \frac{D(u)A_p^2(u)du}{A_p^2(E)P_p(u)}}\right]
\end{eqnarray} 
where $P_p(E)=V_0E+B_p(E)$ and
\begin{equation}
A_p(E)=exp\left[3V_0\int^E\frac{du}{V_0u+B_p(u)}\right]\nonumber
\end{equation} 
\begin{equation}
A_{pp}=8\times 10^{-16} \bar{n}_H\nonumber
\end{equation} 
Then, the general solution of equation (9) is obtained as
\begin{eqnarray}
N_p(x,y,z,E)=\int^{\infty}_{-\infty}dx^{'}\int^{\infty}_{-\infty}dy^{'}\int^{\infty}_{-\infty}dz^{'}\int^{\infty}_EdE^{'}Q_p(x^{'},y^{'},z^{'},E^{'})\nonumber\\
\times G(x,x^{'},y,y^{'},z,z^{'},E,E^{'})
\end{eqnarray} 
Now, considering a source of the form $Q(x^{'},y^{'},z^{'},E^{'})=q(E^{'})\delta(x^{'}-x_0)\delta(y^{'}-y_0)\delta(z^{'}-z_0)$, the differential proton number density at $(x,y,z)$ due to a single source located at $(x_0,y_0,z_0)$ is obtained as
\begin{eqnarray}
N_p(x,x_0,y,y_0,z,z_0,E)=\frac{A_p(E)}{8\pi^{3/2}P_p(E)}\int^{\infty}_E dE^{'}q(E^{'}){\left[\int^{E{'}}_E \frac{D(u)du}{P_p(u)}\right]}^{-1}\nonumber\\
\times{\left[\int^{E{'}}_E \frac{D(u)A_p^2(u)du}{P_p(u)}\right]}^{-1/2}exp\left[A_{pp}\int^E_{E^{'}} \frac{du}{P_p(u)}\right]\nonumber\\
\times exp\left[\frac{-((x_0-x)^2+(y_0-y)^2)}{4\int^{E^{'}}_E \frac{D(u)du}{P_p(u)}}\right]\nonumber\\
\times exp\left[\frac{-\left(3V_0z_0\int^E_{E^{'}}\frac{A_p(u)du}{A_p(E)P_p(u)}-z_0+z\right)^2}{4\int^{E^{'}}_E \frac{D(u)A_p^2(u)du}{A_p^2(E)P_p(u)}}\right]
\end{eqnarray} 
This gives the number density at $(x=0,y=0,z)$ as
\begin{eqnarray}
N_p(x_0,y_0,z,z_0,E)=\frac{A_p(E)}{8\pi^{3/2}P_p(E)}\int^{\infty}_E dE^{'}q(E^{'}){\left[\int^{E{'}}_E \frac{D(u)du}{P_p(u)}\right]}^{-1}\nonumber\\
\times{\left[\int^{E{'}}_E \frac{D(u)A_p^2(u)du}{P_p(u)}\right]}^{-1/2}exp\left[A_{pp}\int^E_{E^{'}} \frac{du}{P_p(u)}\right]\nonumber\\
exp\left[\frac{-(x_0^2+y_0^2)}{4\int^{E^{'}}_E \frac{D(u)du}{P_p(u)}}\right]
exp\left[\frac{-\left(3V_0z_0\int^E_{E^{'}}\frac{A_p(u)du}{A_p(E)P_p(u)}-z_0+z\right)^2}{4\int^{E^{'}}_E \frac{D(u)A_p^2(u)du}{A_p^2(E)P_p(u)}}\right]
\end{eqnarray} 
which in cylindrical co-ordinates $(r_0,\theta_0,z_0)$ becomes
\begin{eqnarray}
N_p(r_0,\theta_0,z,z_0,E)=\frac{A_p(E)}{8\pi^{3/2}P_p(E)}\int^{\infty}_E dE^{'}q(E^{'}){\left[\int^{E{'}}_E \frac{D(u)du}{P_p(u)}\right]}^{-1}\nonumber\\
\times{\left[\int^{E{'}}_E \frac{D(u)A_p^2(u)du}{P_p(u)}\right]}^{-1/2}exp\left[A_{pp}\int^E_{E^{'}} \frac{du}{P_p(u)}\right]\nonumber\\
exp\left[\frac{-r_0^2}{4\int^{E^{'}}_E \frac{D(u)du}{P_p(u)}}\right]
exp\left[\frac{-\left(3V_0z_0\int^E_{E^{'}}\frac{A_p(u)du}{A_p(E)P_p(u)}-z_0+z\right)^2}{4\int^{E^{'}}_E \frac{D(u)A_p^2(u)du}{A_p^2(E)P_p(u)}}\right]
\end{eqnarray} 
For a uniform source distribution in the galactic disk with density $F(r_0,\theta_0,z_0)=k\delta(z_0)$, the number density $N(E)$ at $(0,0,z)$ due to all the sources located in the disk region with radial distances between $r_1$ and $r_2$ from $(0,0,0)$ can be calculated by integrating equation (15) over the galactic region bounded by $r_1$ and $r_2$ as
\begin{eqnarray}
N_p(z,E)=\frac{kA_p(E)}{2\sqrt{\pi}P_p(E)}\int^{\infty}_E dE^{'}q(E^{'}){\left[\int^{E{'}}_E \frac{D(u)A_p^2(u)du}{P_p(u)}\right]}^{-1/2}\nonumber\\
\times exp\left[A_{pp}\int^E_{E^{'}} \frac{du}{P_p(u)}\right]
\left[exp\left(\frac{-r_1^2}{a_p}\right)-exp\left(\frac{-r_2^2}{a_p}\right)\right]\nonumber\\
\times exp\left[\frac{-z^2}{4\int^{E^{'}}_E \frac{D(u)A_p^2(u)du}{A_p^2(E)P_p(u)}}\right]
\end{eqnarray} 
where
\begin{equation}
a_p=4\int^{E^{'}}_E \frac{D(u)du}{P_p(u)}\nonumber
\end{equation} 
This equation represents the variation of the proton density with the galactic height. For a CR source distribution in infinite space, by setting $z=0$ equation (16) represents the CR proton density at any point in the galactic disk due to sources located within the galactic disk region bounded by $r_1$ and $r_2$ from the point. Taking $r_2\rightarrow \infty$, equation (16) gives the proton density at $(0,0,z)$ due to all the sources located in the disk beyond a distance $r_1$ from $(0,0,0)$ as
\begin{eqnarray}
N_p(z,E)=\frac{kA_p(E)}{2\sqrt{\pi}P_p(E)}\int^{\infty}_E dE^{'}q(E^{'}){\left[\int^{E{'}}_E \frac{D(u)A_p^2(u)du}{P_p(u)}\right]}^{-1/2}\nonumber\\
\times exp\left[A_{pp}\int^E_{E^{'}} \frac{du}{P_p(u)}\right]exp\left(\frac{-r_1^2}{a_p}\right)exp\left[\frac{-z^2}{4\int^{E^{'}}_E \frac{D(u)A_p^2(u)du}{A_p^2(E)P_p(u)}}\right]
\end{eqnarray} 
In the case when $r_1\rightarrow 0$, equation (17) gives the number density at any galactic height $z$ due to all the sources present in the galactic disk as
\begin{eqnarray}
N_p(z,E)=\frac{kA_p(E)}{2\sqrt{\pi}P_p(E)}\int^{\infty}_E dE^{'}q(E^{'}){\left[\int^{E{'}}_E \frac{D(u)A_p^2(u)du}{P_p(u)}\right]}^{-1/2}\nonumber\\
\times exp\left[A_{pp}\int^E_{E^{'}} \frac{du}{P_p(u)}\right]exp\left[\frac{-z^2}{4\int^{E^{'}}_E \frac{D(u)A_p^2(u)du}{A_p^2(E)P_p(u)}}\right]
\end{eqnarray} 
\subsection{Propagation of CR electrons in the galaxy}
The form of the transport equation for CR electrons in the galaxy differs from that of the protons only in the absence of the catastrophic loss term as
\begin{eqnarray}
\nabla\cdot \lbrace D(\textbf{r},E)\nabla N_e(\textbf{r},E)-\textbf{V}(\textbf{r})N_e(\textbf{r},E)\rbrace\qquad\qquad\qquad\qquad\qquad\qquad\nonumber\\
+\frac{\partial}{\partial E}\left\lbrace \left( \frac{\nabla\cdot\textbf{V}}{3}E-\beta_e(\textbf{r},E)\right) N_e(\textbf{r},E) \right\rbrace+Q_e(\textbf{r},E)=0\qquad
\end{eqnarray}
Here, $\beta_e(\textbf{r},E)=-B_e(E)$ represents the total electron energy loss rate due to ionization, bremsstrahlung, inverse compton and synchrotron processes. From Atoyan et al. 1995 [29],
\begin{eqnarray}
B_e(E)=3.07\times 10^{-16}\bar{n}_H+1.0\times 10^{-15}\bar{n}_HE+\nonumber\\
1.01\times 10^{-16}(w_{ph}+w_B)E^{2} 
\end{eqnarray} 
in GeV s$^{-1}$. Here, $\bar{n}_H$ is the mean hydrogen atom density in atoms cm$^{-3}$, $w_{ph}=w_{MBR}+w_{NIR}+w_{FIR}$ and $w_{MBR}$, $w_{NIR}$, $w_{FIR}$, $w_B$ are the energy densities of the microwave background radiation, difuse NIR/optical radiation, diffuse FIR radiation and magenetic field respectively in eV cm$^{-3}$. Following the same mathematical calculations given in section 3.2 for protons on the assumption of sources uniformly distributed in the galactic disk, the electron density $N_e(z,E)$ at $(0,0,z)$ due to all the sources located in the galactic region bounded by the radial distances $r_1$ and $r_2$ from $(0,0,0)$ is obtained as 
\begin{eqnarray}
N_e(z,E)=\frac{kA_e(E)}{2\sqrt{\pi}P_e(E)}\int^{\infty}_E dE^{'}q(E^{'}){\left[\int^{E{'}}_E \frac{D(u)A_e^2(u)du}{P_e(u)}\right]}^{-1/2}\nonumber\\
\left[exp\left(\frac{-r_1^2}{a_e}\right)-exp\left(\frac{-r_2^2}{a_e}\right)\right]exp\left[\frac{-z^2}{4\int^{E^{'}}_E \frac{D(u)A_e^2(u)du}{A_e^2(E)P_e(u)}}\right]
\end{eqnarray} 
where $P_e(E)=V_0E+B_e(E)$ and
\begin{equation}
A_e(E)=exp\left[3V_0\int^E\frac{du}{V_0u+B_e(u)}\right]\nonumber
\end{equation} 
\begin{equation}
a_e=4\int^{E^{'}}_E \frac{D(u)du}{P_e(u)}\nonumber
\end{equation} 
The electron density at $(0,0,z)$ due to all the sources located beyond a distance $r_1$ from $(0,0,0)$ in the galactic disk is given by the equation 
\begin{eqnarray}
N_e(z,E)=\frac{kA_e(E)}{2\sqrt{\pi}P_e(E)}\int^{\infty}_E dE^{'}q(E^{'}){\left[\int^{E{'}}_E \frac{D(u)A_e^2(u)du}{P_e(u)}\right]}^{-1/2}\nonumber\\
\times exp\left(\frac{-r_1^2}{a}\right)exp\left[\frac{-z^2}{4\int^{E^{'}}_E \frac{D(u)A_e^2(u)du}{A_e^2(E)P_e(u)}}\right]
\end{eqnarray} 
and the number density at any galactic height $z$ due to all the sources present in the galactic disk is given by
\begin{eqnarray}
N_e(z,E)=\frac{kA_e(E)}{2\sqrt{\pi}P_e(E)}\int^{\infty}_E dE^{'}q(E^{'}){\left[\int^{E{'}}_E \frac{D(u)A_e^2(u)du}{P_e(u)}\right]}^{-1/2}\nonumber\\
\times exp\left[\frac{-z^2}{4\int^{E^{'}}_E \frac{D(u)A_e^2(u)du}{A_e^2(E)P_e(u)}}\right]
\end{eqnarray} 
\begin{figure}[h]
\centering
\includegraphics*[width=0.5\textwidth,angle=270,clip]{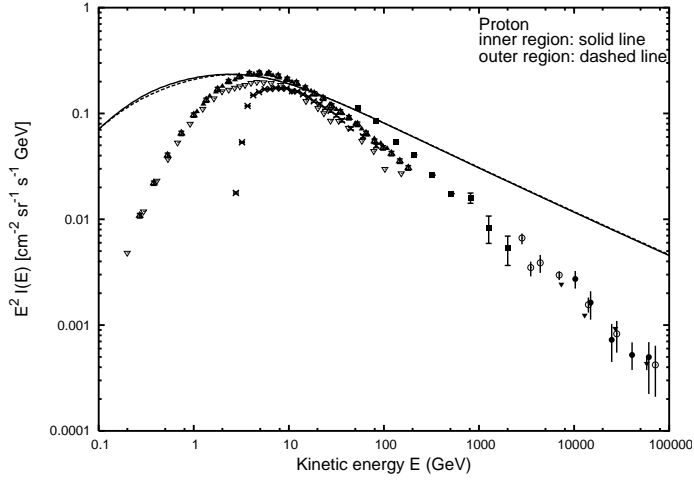}
\caption{\label {fig5} Proton spectra in the inner (solid line) and outer (dashed line) galactic disks $(z=0)$ calculated in the framework of convection-diffusion model using equation (18) for an injection spectrum given by equation (6). The spectra are normalized to the observations at $10 GeV$. The calculation assumes $\bar{n}_H=0.5(0.3)cm^{-3}$ and $\bar{n}_{HII}=0.05(0.03)cm^{-3}$ for inner(outer) galaxy respectively. It can be seen that the two spectra almost overlap each other when normalised to observations at $10 GeV$. Data points : $\blacksquare$ Ryan (1972) [31], $\vartriangle$ Alcaraz 2000 [32], $\times$ Belloti 1999 [33], $\blacktriangle$ Sanuki 2000 [34], $\triangledown$ Boezio 1999 [35], $\circ$ SOKOL [36], $\blacktriangledown$ JACEE [37] and $\bullet$ MUBEE [38].} 
\end{figure}

\begin{figure}[h]
\centering
\includegraphics*[width=0.5\textwidth,angle=270,clip]{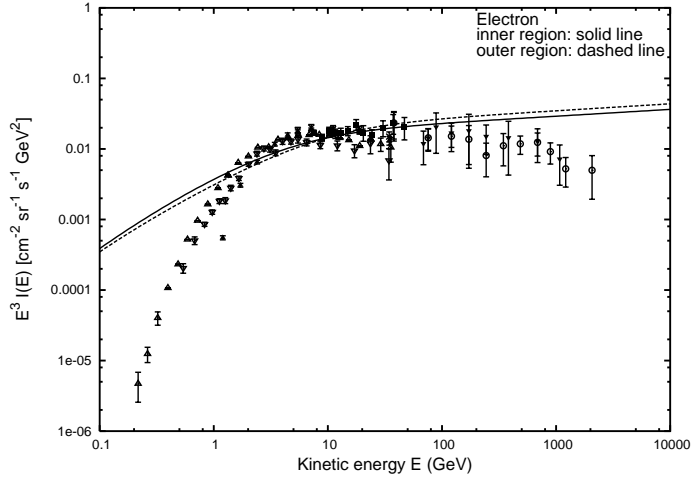}
\caption{\label {fig6} Electron spectra in the inner (solid line) and outer (dashed line) galactic disks $(z=0)$ in the framework of convection-diffusion model calculated using equation (23) normalized to observations at $10 GeV$ for an injection spectrum given by equation (6). The calculation assumes $\bar{n}_H=0.5(0.3)cm^{-3}$, $\bar{n}_{HII}=0.05(0.03)cm^{-3}$, $B=10(6)\mu G$ and $w_{NIR}= 2.0(0.5) eV/cm^3$ for the inner(outer) galaxy. For both the regions $w_{MBR}= 0.25 eV/cm^3$ and $w_{FIR}= 0.2 eV/cm^3$. Data points : $+$ Alcaraz 2000 [39], $\circ$ Boezio 2000 [40], $\times$ Du Vernois 2001 [41], $\vartriangle$ Grimani 2002 [42], $\blacksquare$ Nishimura 1980 [43], $\blacktriangle$ Nishimura 2001 [44], $\square$ Barwick 1998 [45].} 
\end{figure}
\subsection{Calculated particle spectrum compared with observed spectrum}
In the convection-diffusion model, for CR protons and low energy electrons which do not suffer significant radiative losses we can define a characteristic scale length $z_c$ which is basically the height above the galactic plane at which the characteristic diffusive time scale $\sim z^2/D(E)$ equals the convective timescale $V_0^{-1}$ as $z_c=[2D(E)/(3+\alpha/2)V_0]^{½}$ [46,47]. The effect of convection becomes visible only when $z_c<<z_h$, the halo boundary which is taken as infinity in our model. $z_c$ is essentially the boundary between a diffusion dominated region $(z<z_c)$ and a convection dominated region $(z>z_c)$ and in fact the boundary of an effective diffusion halo beyond which the CR density decreases strongly. It plays the same role as the diffusion halo in the pure diffusion models except that it is energy dependent. For the diffusion dominated regime i.e. $z<<z_c$, say at $z = 0$, $N(E)\propto E^{-(\Gamma + \alpha /2)}$. This spectrum is flatter than the one expected from pure diffusion models which can be obtained here by setting $V_0 \rightarrow 0$ in our model. For this, $z_c \rightarrow \infty$ from the above relation which implies $z_c\sim z_h$ and the effect of convection becomes negligible. Then, our model approaches that of the pure diffusion models which gives $N(E)\propto E^{-(\Gamma + \alpha)}$. Similar characteristic scale length can also be defined for high energy electrons which suffer significant radiative losses (with rates given by $-bE^2$ where $b$ is a constant) as $z_d=[4D(E)/((1-\alpha )bE)]^{½}$ [46]. But here, the competitive timescales which define the scale length are the diffusive timescale and the radiative losses timescale. For such high energy electrons, the electron distribution approaches that of the pure diffusion model with radiative losses and is given by $N(E)\propto E^{-(\Gamma + \alpha /2 + 1/2)}$ for $z<<z_d$. 

The spectrum of protons and electrons calculated using equations (18)$\&$(23) respectively for the inner(solid line) and outer(dashed line) galactic disks (i.e at $z=0$) for a uniform source distribution in the galactic disk are shown in Figs. 5$\&$6 together with direct measurements in the solar vicinity. Proton data points are taken from [31-38] and the electron data are from [39-45]. The injection rates of the particles from the sources are such that all the model spectra are normalized to the observed spectra at $10 GeV$ since the solar modulation effects are believed to be negligible for particle energies $\geq 10 GeV$. For protons, it can be seen that the spectrum in the inner and outer regions almost overlap each other when normalised to observations at $10 GeV$. But, for electrons the spectrum in the inner region is found to show a slight steepening compared to that in the outer region. The values of the model parameters used in the calculations are as follows:

The total galactic magnetic field B is found to vary with the galactic radius R [48]. Its value is about 10 $\mu G$ at $R=3$ kpc, and decreases to about 6 $\mu G$ near $R=8.5$ kpc, and then down to 4 $\mu G$ at around $R=17$ kpc. For the present analysis, B is taken as 10(6) $\mu G$ for inner(outer) galaxy. For the mean density of ionised gas in the galaxy $\bar{n}_{HII}=\int_{R_1}^{R_2}\bar{n}_{HII}(R)dR/(R_2-R_1)$ where $\bar{n}_{HII}(R)=\int_{0}^{h}n_{HII}(R,z)dz/h$, the density distribution $n_{HII}(R,z)$ given in [49] is adopted. ($R_1,R_2$) takes the same value as in section 3.1 for calculating $\bar{n}_H$. This gives $\bar{n}_{HII}=0.05$ $cm^{-3}$ in the inner galaxy and $\bar{n}_{HII}=0.03$ $cm^{-3}$ in the outer region. $\bar{n}_H$ is taken as same as in section 3.1. Though the density of 2.7K cosmic microwave background radiation is almost constant in the universe with a value of $w_{MBR}= 0.25 eV/cm^3$, other ISRFs densities like the diffuse NIR/optical radiation and the diffuse FIR radiation fields do not remain constant in the galaxy. $w_{NIR}$ varies from $\sim$ 0.5 eV/cm$^3$ at $R=8.5$ kpc [50] to $\sim$ 2.5 eV/cm$^3$ at around $R=1$ kpc in the galaxy. $w_{FIR}$ varies in the range of $\sim$ 0.2- 0.3 eV/cm$^3$ [51] in the galactic plane. Here, a value of $w_{NIR}= 2.0(0.5) eV/cm^3$ for inner(outer) region and $w_{FIR}= 0.2 eV/cm^3$ for both the regions are adopted. Also, the SN explosion energy is taken as $E_{SN}=10^{51}$ ergs, ejecta mass as $M_{ej}$= $8M_\circledS$, injected particle density and energy in the SN shock as $n_0= 10^{-3}$ cm$^{-3}$ and $1$ KeV respectively, and an upper energy cut-off of $100TeV$ in the source term.

It is found from the calculations that the average proton and electron spectra in both the inner and outer galaxy over-predict the observed data both at lower and higher energies. The over-prediction of the model spectra above the data in the lower energy region may be attributed to the solar modulation effects, but that in the higher energy region (E$\gtrsim 10$ GeV) where the solar modulation effects are less, clearly shows that both the protons and electrons spectra in the galactic disk are flatter than the observed spectra. This deviation of the calculated spectrum from the observed spectrum may be understood as follows: for example, because of the small energy-loss timescale ($t_{loss}$) for high energy electrons compared to diffusive escape timescale, they cannot travel a long distance in the galaxy and only the electrons from nearby sources may contribute to the local spectrum. For diffusion coefficient $D\sim (1-10)\times 10^{28}$ cm$^2$ s$^{-1}$, the range of a 100 GeV electron ($t_{loss}\approx 10^6 yr$) is estimated as $R\approx 2\sqrt {Dt_{loss}}\approx (200-600) pc$. That means only those sources within a distance $d\approx (200-600) pc$ from the Sun can contribute to the local electron spectrum at 100 GeV. Detailed calculations done by Atoyan et al.1995 [29] have shown that the total CR electron flux near the Sun even at energies $1-100$ GeV where the particles do not suffer significant energy losses, calculated assuming a uniform and continuous source distribution in the galactic disk in the framework of \textit{diffusive} (without convection) propagation model, is dominated by particles injected from sources with distances $d\leq 1 kpc$. Therefore, it is quiet necessary to know the electron spectrum generated by local sources together with the local CR propagation parameters, in addition to the galactic averaged spectrum (here considered as the spectrum generated by distant sources) if the observed spectrum is to be understood properly. Detailed discussions about the local CR spectra (both protons and electrons) will be discussed in section 5. 

It should also be mentioned at this point of discussion that in the case of a stationary injection of CRs into the galaxy from sources continuously distributed in the galactic disk, the equilibrium CR (proton and electron) spectra in the disk depends weakly on the actual geometry of the source distribution as long as the CR energy densities are normalized to the same value. Another important point is that away from the galactic plane, the shape of the CR electron spectrum shows a significant variation from that in the disk particularly in the high and low energy regions due to their high radiative and ionization energy losses at these energies i.e. the spectrum is softer by an index of $\triangle \Gamma\sim 0.6$ at a distance of $1kpc$ away from the galactic plane for energies greater than $\sim 100GeV$. But, the proton spectral shape is almost the same as in the disk except in the very low energy region due to their high ionization energy losses operating at these energies. So, in the following section while calculating the diffuse gamma-ray spectrum averaged over a galactic region, such a variation of the CR spectra with distance away from the disk will be taken care.  

\section{Diffuse gamma-ray spectra}
There are three important processes for the production of diffuse non-thermal high energy gamma-rays in the galaxy due to the interaction of galactic CRs with the ISM and ISRF. They are the inverse compton (IC) and bremsstrahlung of electrons and the decay of $\pi^0$-mesons produced in inelastic collisions of CR protons and nuclei with ambient nucleons. 
\subsection{Gamma-ray emissivities} 
The $\pi^0$-mesons decay $\gamma$-rays emissivity is given by [52] and references therein as
\begin{equation}
q_{\gamma}^{\pi^0}(z,E_\gamma)=2\int_{E^{min}_{\pi^0}(E_\gamma)}^{\infty} \frac{q_{\pi^0}(z,E_{\pi^0})}{\sqrt{E_{\pi^0}^2-m_{\pi^0}^2c^4}}dE_{\pi^0}
\end{equation} 
where $E^{min}_{\pi^0}(E_\gamma)=E_\gamma+m_{\pi^0}^2c^4/4E_\gamma$ and $m_{\pi^0}$ is the $\pi^0$-meson rest mass energy. The emissivity of $\pi^0$ mesons is calculated as
\begin{equation}
q_{\pi^0}(z,E_{\pi^0})=4\pi \bar{n}_H\eta\int_{E_p^{min}(E_{\pi^0})}^{\infty}dE_pJ_p(z,E_p)\frac{d\sigma_{\pi^0}(E_{\pi^0},E_p)}{dE_{\pi^0}}
\end{equation} 
where $J_p(z,E_p)\approx(c/4\pi)N_p(z,E_p)$ is the CR proton flux, $d\sigma_{\pi^0}(E_{\pi^0},E_p)/dE_{\pi^0}$ is the differential crossection for a CR proton of kinetic energy $E_p$ to produce a $\pi^0$ with total energy $E_{\pi^0}$, $\eta=1.5$ is the nuclear enhancement factor which takes into account the contribution of heavier nuclei in CRs and ISM to the the $\gamma$-ray emission and $E_p^{min}(E_{\pi^0})$ is the minimum proton kinetic energy that contributes to the production of a pion with energy $E_{\pi^0}$.

The emissivity of IC $\gamma$-rays due to the interactions of CR electrons and the ISRFs using the full Klein-Nishina cross-section is given by [53]
\begin{eqnarray}
q_{\gamma}^{IC}(z,E_\gamma)=8\pi^2r_0^2(m_e^2)\frac{n_{ph}}{E_{ph}}\int_{E_{th}}^{\infty}dE_e\frac{J_e(z,E_e)}{E_e^2}\qquad\qquad\qquad\nonumber\\
\left[2q \ln q+(1+2q)(1-q)+\frac{1}{2}\frac{{(\Gamma q)}^2(1-q)}{1+\Gamma q}\right]
\end{eqnarray} 
where
\begin{equation}
\Gamma =\frac{4E_{ph}E_e}{m_e^2}\nonumber\\
\end{equation} 
\begin{equation}
q=\frac{E_\gamma}{E_e\Gamma (1-E_\gamma/E_e)}\nonumber\\
\end{equation} 
\begin{equation}
E_{th}=\frac{1}{2}\left[E_\gamma+\sqrt{E_\gamma^2+\dfrac{E_\gamma m_e^2}{E_{ph}}}\right]\nonumber\\
\end{equation} 
\begin{equation}
J_e(z,E_e)\approx\frac{c}{4\pi}N_e(z,E_e)\nonumber\\
\end{equation} 
\begin{equation}
n_{ph}=\frac{w_{ph}}{E_{ph}}\nonumber\\
\end{equation} 
$w_{ph}$ and $n_{ph}$ are the energy and number densities of the target photon respectively, $E_{ph}$ is the energy of the target photon, $E_{th}$ is the minimum electron energy that can scatter a target photon of energy $E_{ph}$ to energy $E_\gamma$, $r_0=2.82\times 10^{-13}$ $cm$ is the classical electron radius and $m_e$ is the electron rest mass energy. 

\begin{figure}[h]
\centering
\includegraphics*[width=0.5\textwidth,angle=270,clip]{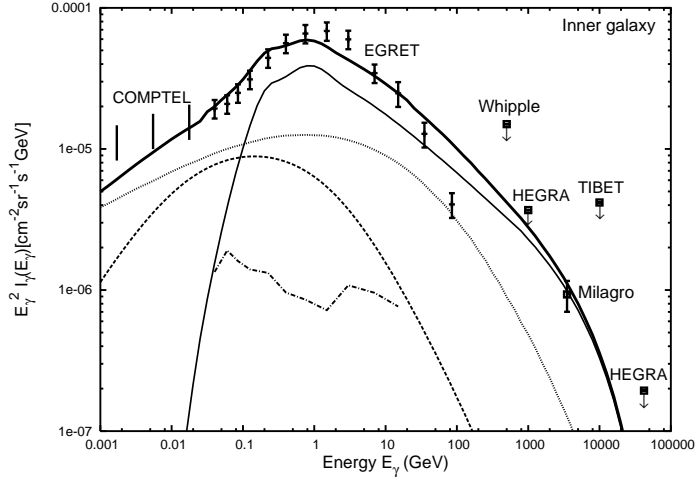}
\caption{\label {fig7} Averaged diffuse $\gamma$-ray spectrum in the inner galaxy ($330^\circ<l<30^\circ$, $|b|<5^\circ$) calculated using the CR proton and electron spectra in the region, assuming $w_p= 1.4 eV/cm^3$, $w_e= 0.07 eV/cm^3$, $n_H= 1 cm^{-3}$, $w_{MBR}= 0.25 eV/cm^3$, $w_{NIR}= 2.0 eV/cm^3$, $w_{FIR}= 0.2 eV/cm^3$, $B=10\mu G$ and $r_d= 15 kpc$. COMPTEL and EGRET data points are as in [9], Milagro [56], Whipple [58], HEGRA [59,60] and TIBET [61]. Model components : $\pi^0$ decay (thin solid line), bremsstrahlung (dashed line), IC (dotted line), extragalactic diffuse $\gamma$- ray background (dot-dashed line) taken from [55] and total (thick solid line).}
\end{figure}

\begin{figure}[h]
\centering
\includegraphics*[width=0.5\textwidth,angle=270,clip]{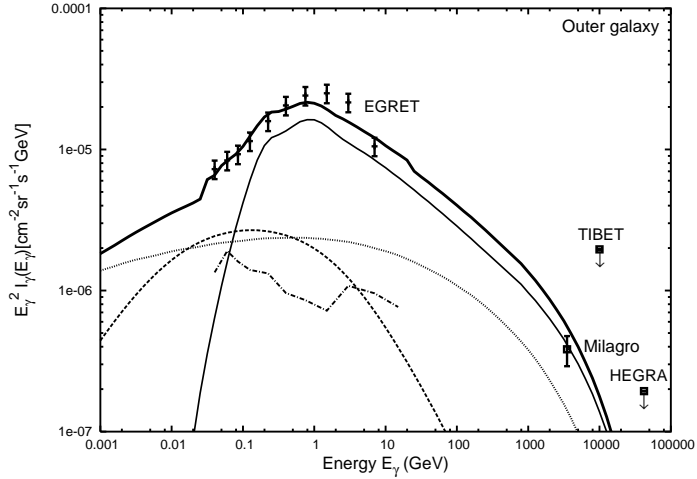}
\caption{\label {fig8} Averaged diffuse $\gamma$-ray spectrum in the outer galaxy ($30^\circ<l<330^\circ$, $|b|<5^\circ$) calculated using the CR proton and electron spectra in the region, assuming $w_p= 1.0 eV/cm^3$, $w_e= 0.05 eV/cm^3$, $n_H= 0.6 cm^{-3}$, $w_{NIR}= 0.5 eV/cm^3$, $B=6\mu G$, $r_d= 9.5 kpc$ and all the other model parameters are same as in Fig.7. Data points: EGRET [9], Milagro [56], HEGRA [60] and TIBET [61].}
\end{figure}

The emissivity of bremsstrahlung $\gamma$-rays due to the interactions of CR electrons and the ISM is given by [53]
\begin{equation}
q_\gamma^{brem}(z,E_\gamma)=4\pi \bar{n}_H\int_{E_{e,min}}^{\infty}J_e(z,E_e)\frac{d\sigma_{brem}(E_e,E_\gamma)}{dE_\gamma}dE_e
\end{equation} 
where
\begin{eqnarray}
\frac{d\sigma_{brem}(E_e,E_\gamma)}{dE_\gamma}=4\alpha r_0^2z^2\frac{1}{E_\gamma}\left[\frac{4}{3}-\frac{4E_\gamma}{3E_e}+\frac{E_\gamma^2}{E_e^2}\right]\nonumber\\
\left\{\ln \left[\frac{2E_e}{m_e}\frac{(E_e-E_\gamma)}{E_\gamma}\right]-\frac{1}{2}\right\}
\end{eqnarray} 
\begin{equation}
E_{e,min}=max[E_\gamma,E_L]\nonumber\\
\end{equation} 
where $\alpha=\frac{1}{137}$ is the fine structure constant, $z$ is the atomic number of the target atom and $E_L$ is the lowest energy in the electron spectrum.
\subsection{Total gamma-ray flux} 
Using the $\gamma$-ray emissivities listed above in section 4.1, the total $\gamma$-ray flux with energy $E_\gamma$ in a given direction denoted by longitude $(l)$ and latitude $(b)$ is calculated as
\begin{equation}
I_\gamma(l,b,E_\gamma)=\frac{1}{4\pi}\int [q_{\gamma}^{\pi^0}(z,E_\gamma)+q_{\gamma}^{IC}(z,E_\gamma)+q_\gamma^{brem}(z,E_\gamma)]dr
\end{equation} 
The integration is over the line of sight distance $r$ measured from the Sun in the direction of $(l,b)$ and $z=r$ $sin$ $b$ denotes the height above the galactic plane. Notice that $I_\gamma(l,b,E_\gamma)$ is independent of $l$ in the present model since both the CR proton and electron spectra are assumed to be constant in the galactic disk $(z=0)$ as discussed in section 3. For calculating the $\pi^0$ decay and bremsstrahlung $\gamma$-rays fluxes, the mean ISM hydrogen atom density is taken as $\bar{n}_H=1 cm^{-3}$ in the inner galaxy and $\bar{n}_H=0.6 cm^{-3}$ in the outer region (as discussed in section 3.1). For IC $\gamma$-ray flux, the target photons considered here are the 2.7 K cosmic microwave background radiation, diffuse NIR/optical radiation and the diffuse FIR radiation fields. Their energy densities are taken as $w_{MBR}=0.25$ eV/cm$^3$ and $w_{FIR}=0.2$ eV/cm$^3$ in both the galactic regions, and $w_{NIR}=2.0(0.5)$ eV/cm$^3$ in the inner(outer) region (see section 3.2). The overall normalization in calculating the $\pi^0$-decay, bremsstrahlung and IC $\gamma$-rays fluxes goes as $\bar{n}_H\times r_d\times w_p$, $\bar{n}_H\times r_d\times w_e$ and $w_{ph}\times r_d\times w_{e}$ respectively where $w_p,w_e$ and $w_{ph}=w_{MBR}+w_{NIR}+w_{FIR}$ are the protons, electrons and the total target photons energy densities respectively and $r_d$ is the line of sight depth of the gamma-ray emission region. The total CR energy density of protons, helium and heavier nuclei in which the main fraction are protons, varies from the locally measured value of $\sim$ 1.0 eV cm$^{-3}$ at sunspot minimum ($\sim$ 0.8 eV cm$^{-3}$ at sunspot maximum) to a value of $\sim$ 1.6 eV cm$^{-3}$ at $\sim$ 60 AU [54]. Also, the CR electron energy density has been found to vary from the local value of 0.05 eV cm$^{-3}$ to 0.2 eV cm$^{-3}$ at $\sim$ 60 AU [54]. Here, for calculating the diffuse $\gamma$- ray flux in both the inner and outer regions of the galaxy, the particle injection rates from the sources are chosen such that the resulting CR proton energy density $w_p= 1.4(1.0) eV cm^{-3}$ and the electron energy density $w_e= 0.07(0.05) eV cm^{-3}$ for the inner(outer) galaxy. For the velocity integrated CO intensities given in [25], the molecular $H_2$ column density $N(H_2)$ in the direction of the inner galaxy varies approximately in the range of $(0.95-3.4)\times 10^{22}cm^{-2}$ and that in the outer galaxy as $\sim(0.28-1.33)\times 10^{22} cm^{-2}$ for a CO-to-H$_2$ conversion factor $X=1.9\times 10^{20} cm^{-2}(K km s^{-1})^{-1}$. Here, we assume a value of $N(H_2)=1.5\times 10^{22}cm^{-2}$ for the inner galaxy and $N(H_2)=0.4\times 10^{22}cm^{-2}$ for the outer galaxy. Therefore, for the assumed molecular $H_2$ averaged number density of $\bar{n}_{H_2}=0.33(0.14)cm^{-3}$ for the inner(outer) galaxy, the mean line of sight depths of the emission region $r_d\simeq 15(9.5) kpc$ in the inner(outer) galaxy. It should be noted that the individual normalizations $\bar{n}_H,w_p,w_e,w_{ph}$ and $r_d$ may be slightly varied from the above mentioned values but the overall normalization should give the same value. 

The diffuse $\gamma$-ray spectrum calculated using the above set of model parameters, averaged over the latitude range $|b|\leq 5^\circ$ are shown in Figs.7$\&$8 for the inner and outer galaxy respectively together with the available experimental data. The EGRET data shown in the plots are taken from [9,55]. The Milagro point in the outer galaxy is obtained from [56] and for the inner galaxy, a conversion factor of $2.43\pm 0.55$ from $(40^\circ <l<100^\circ)$ to $(-40^\circ <l< 40^\circ)$ is adopted [57]. Other data points are upper limits set by Whipple [58], HEGRA [59,60] and TIBET [61] collaborations. In the inner galactic region, the model diffuse gamma-ray spectra obtained in this paper show a good fit to the observed data at almost all the energies except at $\sim 85 GeV$ where the model predicts a much higher flux than the observed data. Though there is a possibility that the CR spectra derived here may not represent the true spectra at around $85 GeV$, it should also be kept in mind that the effective energy range of EGRET is only about $30 MeV-30 GeV$ [1] where the present model shows a good fit to the data. In the outer galaxy also the model spectra nicely explains the observed gamma-ray data, but the results obtained in this region are somewhat less convincing because of the limitation of significant number of data points. The EGRET points extend only upto $7 GeV$ [9] in this region and beyond that we have only the MILAGRO point at $3.5 TeV$ for $(40^\circ <l<100^\circ ,|b|<5^\circ)$ [56] together with the upper limits given by ground-based experiments like TIBET and HEGRA.

According to Hunter et al. 1997 [1], the spectrum of diffuse gamma-ray emission in the EGRET energy range does not vary significantly with longitude or latitude for $|b|\leq 10^\circ$. The overall results obtained in this paper agree quiet well with those results since the high energy CR proton spectrum hardly varies in the galaxy in our model and the $\pi ^0$-decay gamma-rays dominate the total spectrum in the EGRET region. However, the present model is not able to explain the weak evidence ($\sim 3\sigma$) of softening of the gamma-ray spectrum in the outer galaxy compared to the inner galaxy at very low latitudes $|b|\leq 2^\circ$ mentioned in their paper.  

\section{Local cosmic-ray spectra }
It has been seen from the results discussed in the previous sections that the averaged CR spectrum calculated for the inner and outer regions of the galaxy can explain the observed diffuse $\gamma$-ray spectrum effectively well in both the galactic regions. But, one serious problem which requires detail study is why the galactic averaged CR spectrum which fits the diffuse $\gamma$-ray data so well, is flatter than the observed CR spectrum. This section describes a possible explanation for the problem. 

For CR protons in the framework of a steady 3-D convection-diffusion model, it can be seen from Fig.9 that for a standard diffusion coefficient $D_0=0.7\times 10^{28}cm^2 s^{-1}$, convection velocity $V_0=15 km s^{-1} kpc^{-1}$ and source index $\Gamma_p=2.4$, most of the protons below $\sim 100 GeV$ are coming from sources located within a distance of $d\sim 1.5 kpc$ from the Sun. This is due to the fact that the CR diffusive timescale goes as $(t_{diff}\propto 1/D(E))$ with energy. So, high energy CRs take a lesser time to reach us than the low energy CRs. Those protons with energy less than $100 GeV$ may be either convected away from the galactic plane or may suffer significant energy losses due to ionization and coulomb interactions with the ISM (if their energies are less than $\sim 100 MeV$) or due to adiabatic cooling (if their energies are greater than $\sim 100 MeV$) before reaching us from their sources. So, protons below $\sim 100 GeV$ are not able to reach us effectively from distant sources. The present analysis assumes a constant diffusion coefficient throughout the galaxy as well as a continuous and uniform source distribution in the galactic disk injecting CR protons with a power-law source index $\Gamma_p=2.4$. But, this may always not be the case. If the diffusion coeffient in the local region is different from the galactic averaged value, then the scenario may be much different from what is shown in Fig.9. One such scenario is shown in Fig.10 where the diffusion coefficient is taken as $D_0=10^{27}cm^2 s^{-1}$ for region within $1.5 kpc$ from us and $D_0=5\times 10^{28}cm^2 s^{-1}$ for region beyond $1.5kpc$. The total flux (solid line) is splitted into the contributions from local sources within $1.5 kpc$ (dashed line) and distant sources beyond $1.5 kpc$ (dotted line). It shows that the majority of the CR protons upto $\sim 10 TeV$ are liberated mainly by sources located within $1.5 kpc$ provided the diffusion coefficient in the local region is allowed to differ appreciably from the galactic averaged value. In fact, studies assuming a localized nature of the galactic CRs requires the value of the diffusion coefficient to be $D_0\sim (10^{26}-10^{27})cm^2 s^{-1}$ considering the observed CR anisotropy [62]. So, from now on, it will be assumed that such a variation in the diffusion coefficient exists in the galaxy and so the observed CRs are contributed mainly by local sources. 

For CR electrons, as already mentioned in section 3.4, considering the diffusive escape and energy loss timescales, the majority of the locally observed electrons are mainly generated by sources located within a distance of $d\leq$1 kpc from the Sun (Atoyan et al.1995 [29]). Though the authors had discussed the problem in the framework of a \textit{purely diffusive} model without convection, the result is almost same in the \textit{convection-diffusion} model as well. So, without going into much detail, the observed CR electrons will be asumed to be contributed mainly by sources located within $1.5 kpc$ as in the case of protons. It can be noted that for any acceptable value of $D_0\lesssim 10^{29}cm^2 s^{-1}$, the sources of the observed CR electrons are mainly localized. It is because of the fact that in addition to the diffusive escape timescale, various other timescales like the ionization, bremsstrahlung, inverse compton,synchrotron, adiabatic cooling and convective timescales govern the CR electron spectrum.

\begin{figure}[h]
\centering
\includegraphics*[width=0.5\textwidth,angle=270,clip]{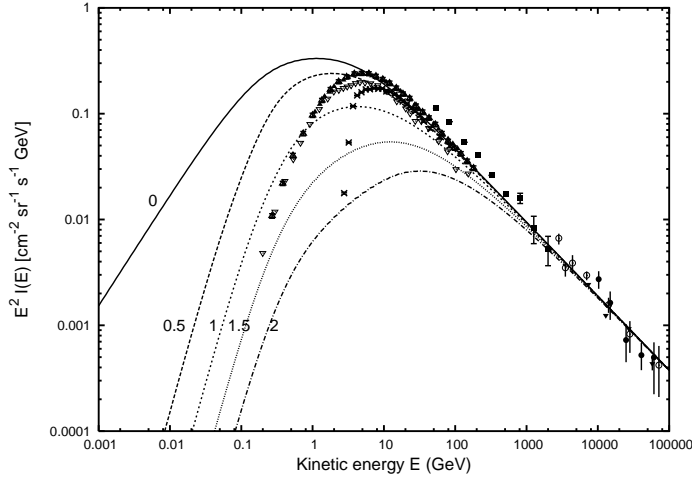}
\caption{\label {fig9} Decomposed CR proton spectrum near the solar region in the framework of a convection-diffusion model for $D_0=0.7\times 10^{28}cm^2 s^{-1}$, $\alpha=0.6$, $V_0=15 km s^{-1} kpc^{-1}$ and source spectral index $\Gamma_p=2.4$. The solid curve (marked as 0) is the total flux decomposed into the contributions from sources located at distances $d\geq (0.5,1.0,1.5,2.0)kpc$ shown by the curves marked as $(0.5,1,1.5,2)$ respectively. The total spectrum is normalized to the observed flux at $10 GeV$ and the calculation assumes $\bar{n}_{H}=1.11 cm^{-3}$ and $\bar{n}_{HII}=0.022 cm^{-3}$. Data points are same as in Fig.5.}
\end{figure}
\begin{figure}[h]
\centering
\includegraphics*[width=0.5\textwidth,angle=270,clip]{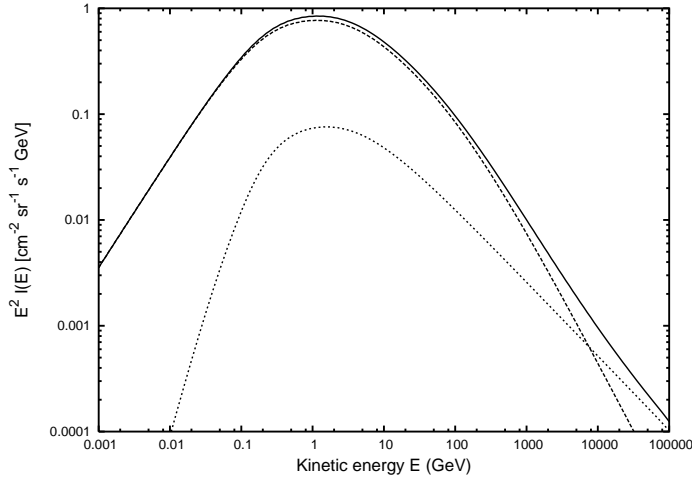}
\caption{\label {fig10} Unnormalized CR proton spectrum calculated using a convection-diffusion model assuming diffusion coefficient $D_0=10^{27}cm^2 s^{-1}$ for the local region (i.e $d<1.5 kpc$), $D_0=5\times 10^{28}cm^2 s^{-1}$ for the region beyond $1.5 kpc$ and convection velocity $V_0=15 km s^{-1} kpc^{-1}$. Here, $\alpha=0.6$ and $\Gamma_p=2.4$. It can be seen from the plot that the total spectrum (solid line) upto $\sim 10 TeV$ is contributed mainly by local sources (dashed line) located within $1.5 kpc$. The contribution from the distant sources located beyond $1.5 kpc$ is represented by the dotted line.}
\end{figure}

Now, to model the observed CR proton and electron spectra considering only the sources located within $1.5 kpc$ from us, we will proceed with the usual steps followed in this paper as described in sections 2$\&$3, i.e. finding the SNR age distribution (here for the local region), then use it to calculate the source spectrum and then propagate the CRs in the galaxy. From a compilation of several parameters of 231 all-type SNRs in the galaxy given in Table 1 of Jian et al. (2005) [12], it is found that there are only 10 S-type SNRs which are located within a distance $d<1.5 kpc$ from the Sun. Out of these, 8 SNRs have known ages. The $l,b,d$ and age($t$) parameters of these 8 SNRs are listed down in Table 2. Column 5 lists the $log_{10}t$ of each SNR. A gaussian fitting on the distribution of these SNRs ages using equation (2) gives the fitted parameters as : $a_0=0.415\pm 0.236, a_1=4.153\pm 0.124$ and $a_2=0.278\pm 0.142$. Using this age distribution for the nearby SNRs, the CR source spectrum is obtained using equation (6) and is found to be $\Gamma_p=2.24\pm 0.12$ for protons with $E>>M_p$ and $\Gamma_e=2.27\pm 0.08$ for electrons. The large error bar on the spectral indices are due to the small sample size (only 8 SNRs with known ages) involve in the analysis. After propagating the CRs in the galaxy, the proton and electron fluxes at the solar position can be obtained using equations (16) and (21) respectively by taking $r_1=0$ and $r_2=1.5 kpc$. The energy spectrum thus calculated are 
\begin{table}   
\caption{Parameters of 8 S-type SNRs with $d<1.5 kpc$ :}
\centering
\begin{tabular}{|c|c|c|c|c|} 
\hline
$l$ & $b$ & $d$ & Age($t$) & $log_{10}t$\\
( $^\circ$ ) & ( $^\circ$ ) & (kpc) & (yrs) & \\
\hline
65.3   &    5.7  &     1.0  &   14000  &  4.146\\
73.9   &    0.9  &     1.3  &   10000  &  4.000\\
74.0   &   -8.5  &     0.4  &   14000  &  4.146\\
89.0   &    4.7  &     0.8  &   19000  &  4.279\\
114.3  &    0.3  &     0.7  &   41000  &  4.613\\
119.5  &   10.2  &     1.4  &   24500  &  4.389\\
160.9  &    2.6  &     1.0  &   7700   &  3.886\\
299.2  &   -2.9  &     0.5  &   5000   &  3.699\\
\hline
\end{tabular}
\end{table}
shown in Figs. 11$\&$12 for CR protons and electrons respectively. The calculation assumes a continuous and uniform source distribution, a diffusion coefficient $D_0=0.3\times 10^{27}cm^2 s^{-1}$ and a convection velocity of $V_0=15 km s^{-1} kpc^{-1}$ in the local galactic region. The solid line represents the model spectrum calculated for a sharp energy cut-off at $100 TeV$. For electrons, the model spectra calculated for an exponential energy cut-off at $5 TeV$ (dotted line) and at $10 TeV$ (dashed line) are also shown. All the model spectra are normalized to the observed flux at $10 GeV$. The over-prediction of the model spectra above the observed spectra below $E\lesssim 10$ GeV can be due to the presence of solar modulation effects. In the case of protons, the slight deficit of the model spectrum from the observed spectrum for energies greater than $\sim 10 TeV$ shows the required contribution of the distant sources as expected as discussed above (see Fig.10) in this section, for the allowed variation of the local diffusion coefficient from the typical galactic value. The model parameters used in the calculations are chosen at $(R,z)=(8.5,0) kpc$ as $\bar{n}_H=n_{HI}+2n_{H_2}=1.11$ cm$^{-3}$ where $n_{HI}$ is taken from [24] and $n_{H_2}$ from [25], $\bar{n}_{HII}=0.022$ cm$^{-3}$ [49], $w_{MBR}=0.25$ eV cm$^{-3}$, $w_{NIR}=0.5$ eV cm$^{-3}$ [50], $w_{FIR}=0.2$ eV cm$^{-3}$ [51], SN explosion energy $E_{SN}=10^{51}$ ergs and ejecta mass $M_{ej}$= $8M_\circledS$.

\begin{figure}[h]
\centering
\includegraphics*[width=0.5\textwidth,angle=270,clip]{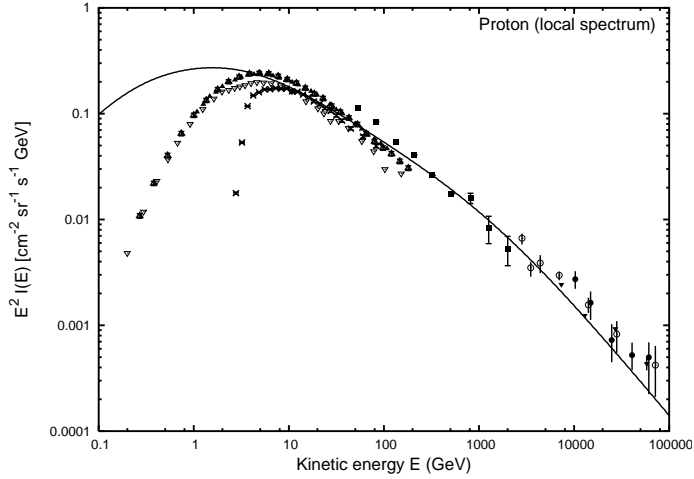}
\caption{\label {fig11} CR proton spectrum calculated in the convection-diffusion model by considering only the sources located within $1.5 kpc$ from the Sun assuming $D_0=0.3\times 10^{27}cm^2 s^{-1}$, $\alpha=0.6$, $V_0=15 km s^{-1} kpc^{-1}$, $\bar{n}_{H}=1.11$ cm$^{-3}$ and $\bar{n}_{HII}=0.022$ cm$^{-3}$. The source spectrum is calculated as described in section 5 and takes a power-law spectral index of approximately $\Gamma_p=2.24\pm 0.12$ for $E>>M_p$ with a sharp energy cut-off at $100 TeV$. The solid line is the calculated model spectrum normalized to observations at $10 GeV$. Data points are same as in Fig.5. The slight deficit of the model spectrum from the observed spectrum for $E\gtrsim 10 TeV$ shows the required contribution of the distant sources in the high energy region (see section 5).}
\end{figure}

\begin{figure}[h]
\centering
\includegraphics*[width=0.5\textwidth,angle=270,clip]{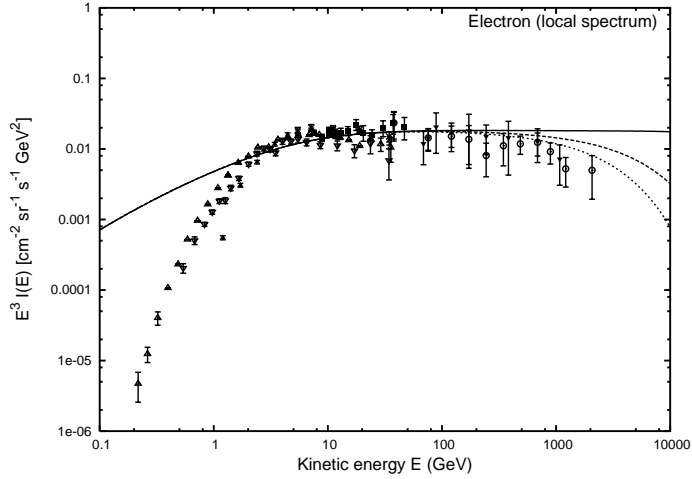}
\caption{\label {fig12} CR electron spectrum calculated in the framework of a convection-diffusion model by considering only the sources located within $1.5 kpc$ from the Sun. The calculated source spectrum adopted here has a power-law index of approximately $\Gamma_e=2.27\pm 0.08$ as described in section 5. All the model parameters are same as in Fig.11. The solid line is the calculated model spectrum for sources with a sharp energy cut-off at $100 TeV$, dotted and dashed lines are for sources with an exponential energy cut-off at $5 TeV$ and $10 TeV$ respectively. All the model spectra are normalized to observations at $10 GeV$. Data points are the same as in Fig.6.}
\end{figure}
\section{Results and discussions }
The model presented in this paper considers SNRs as the main sources of galactic CRs. Considering the SNR age distribution in the galaxy, the model involves the calculation of averaged CR source spectrum both in the inner and outer regions of the galaxy independent of the diffuse $\gamma$-ray data. Earlier attempts to explain the $``GeV$ $excess"$ either assumed an \textit{adhoc} source spectrum optimized on the $\gamma$-ray data or a spectrum obtained from radio observations of SNRs which, as already mentioned, does not have an SNR age dependence (which actually should have according to shock accleration theory). After knowing the source spectra, CRs are allowed to propagate in the galaxy using a steady 3-D convection-diffusion model assuming a uniform and continuous source distribution in the galactic disk. It is found that the resulting CR spectra is able to fit the observed diffuse $\gamma$-ray spectrum quiet well both in the inner and outer regions of the galaxy. 

The present model assumes both the leptonic and nucleonic origin of diffuse gamma-rays in the galaxy like the models discussed in most of the earlier works. But, getting into detail one can notice that the results obtained here differ from the earlier results in some respects. In the inner galaxy at energies around $100MeV-100GeV$, both the $\pi^0$-decay gamma-ray component and the IC component are important to explain the diffuse gamma-ray spectrum and above $\sim 500 GeV$ the $\pi^0$-decay component mainly dominates the total spectrum, unlike in earlier models where enough freedom is given to the $\pi^0$ component such that it alone can possibly explain the GeV energy region and the IC component dominates the TeV region. In the present model, the normalization to the $\pi^0$-decay component is mainly constrained by the flux in the TeV region whereas in the earlier models, it is mainly constrained by the flux in $100MeV-10GeV$ region. Here, fitting the gamma-ray spectrum in the range of $100MeV-10GeV$ by $\pi^0$-decay component of radiation alone will predict a higher flux than the observed data in the TeV range and so the possibility of explaining the $``GeV$ $excess"$ by the $\pi^0$-decay gamma-rays alone is ruled out unless the proton spectrum is too steep with $(\Gamma _p \gtrsim 2.6)$ which again is undesirable since it gives a higher gamma flux in the $100MeV-1GeV$ region. In the present model, the underprediction of the model spectrum below the observed gamma-ray flux in $100MeV-10GeV$ energy region when the $\pi^0$ component is constrained by the observed flux at $3.5 TeV$ is compensated mainly by the IC component. Also, one very important result that can be concluded is that the present model demands the CR proton spectrum to have an exponential cut-off at $\sim 100 TeV$ so that the model spectrum does not exceed the observed flux in the TeV region. This value of proton cut-off energy agrees well with the predictions of shock acceleration in supernova remnants [e.g. 10]. In addition to this, the studies presented here is also able to explain the steepening of the observed CR protons and electrons spectra from the galactic averaged spectra provided the diffusion coefficient $D_0$ in the local region is significantly less than the galactic averaged value. In short, the important results of this paper can be listed as :

(i) In the inner region of the galaxy, the averaged CR proton source spectrum can be represented by a broken power-law in kinetic energy (E) with index $\Gamma _p\simeq$ 2.12$\pm$0.01 for $E>>M_p$ and 1.66$\pm$0.01 for $E<<M_p$ and that of electron by a power-law index $\Gamma _e\simeq$ 2.126$\pm$0.004 for $E>>M_e$.

(ii) In the outer region of the galaxy, the source spectral indices for proton goes as approximately 2.11$\pm$0.01 for $E>>M_p$ and 1.66$\pm$0.01 for $E<<M_p$, and that of electron as 2.124$\pm$0.004 for $E>>M_e$.

(iii) The averaged diffuse gamma-ray spectrum calculated using the galactic CR spectra  explains the observed diffuse $\gamma$-ray spectrum in both the inner and outer regions of the galaxy quiet well. However, the weak evidence ($\sim 3\sigma$) of softening of the gamma-ray spectrum in the outer galaxy compared to the inner galaxy at $|b|\leq 2^\circ$ [56] is still unexplained by the present model.

(iv) The possible explaination of $``GeV$ $excess"$ by the $\pi^0$-decay gamma-rays alone particularly in the inner galaxy is ruled out unless the proton spectrum is too steep with $(\Gamma _p \gtrsim 2.6)$ which again is undesirable since it predicts a higher gamma flux in the $100MeV-1GeV$ region.

(v) Using a gaussian function fitted to the age distribution of only 8 S-type SNRs with known ages located within a distance of $1.5 kpc$ from the Sun, the source spectra in the local region are calculated as $\Gamma_p\simeq 2.24\pm 0.12$ for protons with $E>>M_p$ and $\Gamma_e\simeq 2.27\pm 0.08$ for CR electrons. 

(vi) These local source spectra easily explain the steepening of the observed CR proton and electron spectra from the galactic averaged spectra provided the diffusion coefficient $D_0$ is much less than the galactic averaged value. 

(vii) The present studies support the SNR origin of galactic CRs.
 
Though the model presented here is able to explain the $``GeV$ $excess"$ problem quiet well and at the same time maintaining a local CR spectrum which fits the observed CR data, the results given in this paper involve some uncertainty. Firstly, the errors involve in the estimation of SNR ages (known ages) are quiet large and only for a few SNRs, the ages have a good accuracy. Secondly, in the present analysis the contribution to the galactic CRs from Plerion (P-) type and Composite (C-) type SNRs (which make a total of $\sim 23\%$ of total galactic SNRs) are not considered. Lastly, the large error on the local CR source spectral indices are due to small sample size in the local region (here only 8 SNRs) and can be reduced only by using a statistically large sample. In fact, the number of SNRs detected is quiet less than the total number of SNRs expected. The total number of SNRs in the galaxy can be estimated as $N=\nu T$ where $\nu$ is the galactic supernova explosion rate and $T$ is the average lifetime of an SNR. For $\nu=1/30$ yrs$^{-1}$ and an assumed range of lifetime $T=10^4-10^5$ yrs, $N\sim (333-3333)$. This number strongly depends on the assumed lifetime and for the average value, $T=5\times 10^4$ yrs, $N\sim 1667$. But, a total of only 231 SNRs are given in Green's catalog [11], i.e. only $\sim14\%$ of the expected total number of SNRs have been detected in radio. So, roughly speaking a total of $\sim 60$ SNRs are expected within $d< 1.5$kpc if the SNRs are assumed to be uniformly distributed in the galaxy.

In future work, studies considering the contribution of P- type and C- type SNRs and also the diffuse emission in the low energy region will be addressed in detail.
\section{Acknowledgements}
The author wish to thank A W Strong for providing the EGRET data and also the anonymous referee for his very helpful and important comments.


\begin{thebibliography}{}
\bibitem{}
Hunter, S. D., et al. 1997, ApJ 481, 205
\bibitem{}
Mori, M. 1997, ApJ, 478, 225
\bibitem{}
Simpson, J. A., 1983, in \textit{Composition and origin of Cosmic Rays}, edited by M. M. Shapiro (Reidel, Dordrecht), p. 1.
\bibitem{}
Porter, T. A.,$\&$Protheroe, R. J. 1997, J. Phys. G: Nucl. Part. Phys.,23, 1765
\bibitem{}
Pohl, M.,$\&$Esposito, J.A. 1998, ApJ, 507,327
\bibitem{}
Aharonian, F. A., $\&$ Atoyan, A. M. 2000, A$\&$A 362, 937
\bibitem{}
Berezhko, E. G.,$\&$ V$\ddot{o}$lk, H. J. 2000, ApJ, 540, 923
\bibitem{}
Berezhko, E. G.,$\&$ V$\ddot{o}$lk, H. J. 2004, ApJ, 611, 12
\bibitem{}
Strong, A. W., Moskalenko, I. V., $\&$ Reimer, O. 2004a, ApJ 613, 962
\bibitem{}
Lagage, P. O. $\&$ Cesarsky, C. J., 1983. Astr.Astrophys. 125, 249
\bibitem{}
Green D. A.,1996, Catalogue of SNR University of Cambridge
\bibitem{}
Jian-Wen Xu, Xi-Zhen Zhang $\&$ Jin-Lin Han, chin. J. Astron. Astrophys. Vol. 5 (2005), No. 2, 165
\bibitem{}
Moraal, H., $\&$Axford, W. I. 1983, A$\&$A 125, 204
\bibitem{}
Sturner, S.J. et al. 1997, ApJ, 490, 619
\bibitem{}
Kodaira, K. 1974, PASJ, 26, 255
\bibitem{}
Leahy, D. A., $\&$ Xinji, W. 1989, PASP, 101, 607
\bibitem{}
Berezhko, E. G.,$\&$ V$\ddot{o}$lk, H. J. 1997, Astropart. Phys., 7,183
\bibitem{}
Bell A. R., 1978, MNRAS 182, 147
\bibitem{}
Blandford, R. D., Ostriker, J. P., 1978, Astrophys. J. Letters 221, 229
\bibitem{}
Blandford, R. D., Eichler D. 1987, Phys. Rept. 154, 1
\bibitem{}
Berezhko, E. G., Yelshin V.K., Ksenofontov L.T. 1994, Astropart. Phys. 2, 215
\bibitem{}
Berezhko, E. G., Ksenofontov L.T., Yelshin V.K., 1995, Nuclear. Phys. B Proc.Suppl. 39, 171
\bibitem{}
Kang, H. $\&$ Jones, T. W.,1991, MNRAS 249, 439
\bibitem{}
Gordon, M. A., $\&$ Burton, W. B. 1976, ApJ, 208, 346; Cox, P., Kr$\ddot{u}$gel, E., $\&$ Mezger, P.G. 1986, A$\&$A, 155, 380
\bibitem{}
Bronfman, L., et al. 1988, ApJ, 324, 248
\bibitem{}
Sanders, D. B., Solomon, P. M., and Scoville, N. Z. 1984, ApJ 276, 182
\bibitem{}
Bloemen, J. B. G. M., et al. 1986, Astr. Ap., 154, 25
\bibitem{}
Strong, A. W., $\&$ Mattox, J. R. 1996, A$\&$A, 308, L21
\bibitem{}
Atoyan, A. M, Aharonian, F. A. $\&$ V$\ddot{o}$lk, H. J., 1995, Phys. Rev. D 52 3265
\bibitem{}
Mannheim, K., Schlickeiser, R., 1994, Astr.Astrophys. 286, 983
\bibitem{}
Ryan, M. J., Ormes, J. F. $\&$ Balasubrahmanyam, V. K., Phys. Rev. Letters 28 (1972) 985 $\&$ E1497
\bibitem{}
Alcaraz, J. et al. Physics Letters B 490 (2000a) 27
\bibitem{}
Bellotti, R. et al. Phys. Rev. D60 (1999) 052002
\bibitem{}
Sanuki, T. et al. 2000 astro/ph-0002481
\bibitem{}
Boezio, M. et al. 1999, ApJ 518, 457
\bibitem{}
Ivanenko I P et al. 1993 ICRC(calgary) vol 2, p 17
\bibitem{}
Asakimori K et al. 1998 ApJ 502 278
\bibitem{}
Zatsepin V I et al. 1993 ICRC(calgary) vol 2, p 13
\bibitem{}
Alcaraz, J. et al. Physics Letters B 484 (2000b) 10
\bibitem{}
Boezio, M. et al., 2000, ApJ 532, 653
\bibitem{}
Du Vernois, M. A., et al. 2001, ApJ, 559, 296
\bibitem{}
Grimani, C., et al. 2002, A$\&$A, 392, 287
\bibitem{}
Nishimura, J. et al, 1980, ApJ 238, 394
\bibitem{}
Nishimura, J. et al, 2001, Adv.Space Res Vol 26, No 11, pp 1827 (2000)
\bibitem{}
Barwick, S. W., et al. 1998, ApJ 498, 779
\bibitem{}
Lerche I., Schlickeiser R., 1982, A$\&$A 107, 148
\bibitem{}
Bloemen J.B.G.M., Dogiel V.A., Dorman V.L., Ptuskin V.S., 1993, $A\&$A 267, 372
\bibitem{}
Beck, R., 2001, Space Sci. Reveiws 99, 243-260 
\bibitem{}
Cordes, J.., et al. 1991, Nature, 354, 121
\bibitem{}
Mathis J. S., Mezger P. G., Panagia N., 1983, A$\&$A 128, 212
\bibitem{}
Chi X., Wolfendale A. W., 1991, J. Phys. G 17, 987
\bibitem{}
Dermer, C., 1986, A$\&$A 157, 223
\bibitem{}
Blumenthal, G. R. $\&$ Gould, R. J., 1970, Rev. Mod. Phys. 42, 237 
\bibitem{}
Webber, W. R., 1998, ApJ 506, 329
\bibitem{}
Strong, A.W., Moskalenko, I. V., $\&$ Reimer, O. 2004b, astro-ph/ 0405441
\bibitem{}
Gus Sinnis, 2005, 29th ICRC, Pune, OG.2.1
\bibitem{}
A.D. Erlykin $\&$ A.W. Wolfendale, 2005, 29th ICRC, Pune, OG.2.1
\bibitem{}
LeBohec, S., et al., 2000, astro-ph/0003265
\bibitem{}
Aharonian, F.A., et al., 2001, A$\&$A, 375, 1008
\bibitem{}
Horns, D. $\&$ Schmele, D., 1999, Proc. 26th ICRC, Salt Lake City, OG 3.2.24
\bibitem{}
Amenomori, M., et al., 1997, Proc. 25th ICRC, Durban, vol. 3, 117
\bibitem{}
J. F. Ormes, in \textit{Proceedings of the $18^{th}$ international Cosmic Ray Conference}, Bangalore, 1983, Vol. 2, p. 187 
\end{thebibliography}
\end{document}